\documentclass[12pt]{article}

\newcommand{\nc}{\newcommand}

\usepackage{hyperref}

\nc{\hepref}[1]{\href{http://arXiv.org/abs/#1.html}{#1}}
\nc{\kekref}[1]{\href{http://www-lib.kek.jp/cgi-bin/img_index?#1}{KEK-19#1}}

\usepackage{mcite}
\usepackage{graphicx}
\usepackage{amsfonts}
\usepackage{amsmath}

\nc{\lspace}{\;\;\;\;\;\;\;\;\;\;}  \nc{\llspace}{\lspace \lspace}

\nc{\beq}{\begin{equation}}   \nc{\eeq}{\end{equation}}
\nc{\bea}{\begin{eqnarray}}   \nc{\eea}{\end{eqnarray}}
\nc{\bi}{\begin{itemize}}    \nc{\ei}{\end{itemize}}
\nc{\be}{\begin{enumerate}}  \nc{\ee}{\end{enumerate}}
\nc{\bc}{\begin{center}}     \nc{\ec}{\end{center}}
\nc{\ba}[1]{\begin{array}{#1}}     \nc{\ea}{\end{array}}
\nc{\macierz}[2]{\left(\begin{array}{#1} #2 \end{array}\right)}

\nc{\vts}{\mkern1mu}
\nc{\ri}{\mathrm{i}\vts}
\nc{\sw}{\sin\theta_W}
\nc{\cw}{\cos\theta_W}

\nc{\tb}{\tan\beta}

\nc{\tr}{\mathrm{Tr}}

\nc{\dslash}[1]{\slash\!\!\!#1}

\nc{\CPV}{CPV}

\nc{\Refrmfl}{\Re(f_2^R-\bar{f}_2^L)}
\nc{\ReDZ}{\Re D_Z}
\nc{\ReDgamma}{\Re D_{\gamma}}
\nc{\ggttbar}{g_{\gamma \ttbar}}
\nc{\gZttbar}{g_{Z \ttbar}}
\nc{\gWtb}{g_{Wtb}}

\nc{\nnr}{\nonumber}

\nc{\jpXchecked}{{\Large Formulas X-checked\newline}}

\nc{\jpfigure}[4]{%
  \begin{figure}%
  \begin{center}%
  \includegraphics[#1]{#2}%
  \end{center}%
  \caption{#3}%
  \label{#4}%
  \end{figure}
}%

\nc{\loopC}[2]{{\mathbf{C}}_{#1}^{#2}}

\nc{\fdiag}{0}
\nc{\bg}{B. Grzadkowski}
\nc{\BG}{Bohdan Grzadkowski}
\nc{\lsp}{\;\;\;\;\;\;\;\;}
\nc{\non}{\nonumber}
\nc{\lumun}{\;{\hbox {fb}^{-1}}{\hbox {y}^{-1}}}
\nc{\hc}{\hbox {h.c.}}
\nc{\re}{\hbox {Re}}
\nc{\im}{\hbox {Im}}
\nc{\etal}{\hbox{et al.}}
\nc{\pbarn}{\;\hbox {pb}}
\nc{\ra} {\rightarrow}
\nc{\ctw}{\cos\theta_W} 
\nc{\stw}{\sin\theta_W}
\nc{\ctwsq}{\cos^2\theta_W}        
\nc{\stwsq}{\sin^2\theta_W}
\nc{\ttbar}{t\bar{t}}
\nc{\bbbar}{b\bar{b}}
\nc{\tanb} {\tan \beta}
\nc{\twbdec} {t\rightarrow W^+ b}
\nc{\tbwbdec} {\bar{t} \rightarrow W^- \bar{b}}
\nc{\hprod} {e^+e^- \ra Z^\ast \ra h Z}
\nc{\epem} {e^+e^-}
\nc{\wpwm} {W^+W^-}
\nc{\tbar} {\bar{t}}
\nc{\bbar} {\bar{b}}
\nc{\wpp} {W^+}
\nc{\mt}{m_t}
\nc{\mb}{m_b}
\nc{\mts}{m_t^2}
\nc{\mw} {m_W}
\nc{\mws} {m_W^2}
\nc{\mz} {m_Z}
\nc{\mzs} {m_Z^2}
\nc{\mh} {m_h}
\nc{\mhs} {m_h^2}
\nc{\ma} {m_A}
\nc{\mas} {m_A^2}
\nc{\hdec}{h \ra t\bar{t}}
\nc{\ttbardec}{\ttbar \ra W^+W^-\bbbar}
\nc{\po}{\Phi_1}
\nc{\pht}{\Phi_2}
\nc{\phtd}{\Phi_2^\dagger}
\nc{\phtt}{{\tilde{\Phi}}_2}
\nc{\popo}{\po^\dagger\po}
\nc{\phtpt}{\pht^\dagger\pht}
\nc{\popt}{\po^\dagger\pht}
\nc{\phtpo}{\pht^\dagger\po}
\nc{\sq}{\sqrt{2}}
\nc{\nsd} {N_{SD}}
\nc{\ntt} {N_{tt}}
\nc{\vs}{\vspace{2mm}}
\nc{\sty}{\hat{S}^t_1} \nc{\pty}{\hat{P}^t_1}
\nc{\sts}{(\sty)^2}      \nc{\pts}{(\pty)^2}
\nc{\yts}{\sts+\pts}
\nc{\sby}{\hat{S}^b_1} \nc{\pby}{\hat{P}^b_1}
\nc{\sbs}{(\sby)^2}      \nc{\pbs}{(\pby)^2}
\nc{\ybs}{\sbs+\pbs}
\nc{\eettz}{\epem \rta \ttbar Z}
\nc{\barx}{\bar{x}}
\nc{\bb}{\stackrel{{\scriptscriptstyle (-)}}{b}}

\nc{\et}{\tilde{e}}
\nc{\ft}{\tilde{f}}
\nc{\gt}{\tilde{g}}
\nc{\hti}{\tilde{h}}

\def\sb{s_\beta}
\def\cb{c_\beta}

\def\lsim{\mathrel{\raise.3ex\hbox{$<$\kern-.75em\lower1ex\hbox{$\sim$}}}}
\def\gsim{\mathrel{\raise.3ex\hbox{$>$\kern-.75em\lower1ex\hbox{$\sim$}}}}

\def\mev{\,{\rm MeV}}
\def\gev{\,{\rm GeV}}

\def\rta{\rightarrow}

\def\dps{\displaystyle}
\def\thf{\theta_{\!\scriptscriptstyle{f}}}
\def\sst#1{\scriptscriptstyle{#1}}
\def\ssf{\scriptscriptstyle{f}}

\begin{document}
\renewcommand{\thepage}{-- \arabic{page} --}

%
\font\fortssbx=cmssbx10 scaled \magstep2
\medskip
\begin{flushright}
$\vcenter{
\hbox{\bf IFT-12/2001}
\hbox{March, 2001}
}$
\end{flushright}
\vspace*{1.5cm}
\renewcommand{\thefootnote}{\alph{footnote})}
\begin{center}
{\Large{\bf CP Violation in Higgs-Boson Interactions}}$^{\:}$\footnote{Talk at the 
Cracow Epiphany Conference on b Physics and CP Violation, 5-7 January 2001, Cracow, Poland}\\
\rm
\vspace*{1cm}

{\bf \BG}$^{\:}$\footnote{E-mail: \tt bohdang@fuw.edu.pl}  and
{\bf Jacek Pliszka}$^{\:}$\footnote{E-mail:\tt pliszka@fuw.edu.pl} 

\vspace*{1cm}
{\em Institute of Theoretical Physics, Warsaw University, Warsaw, Poland}\\

\vspace*{2cm}

{\bf Abstract}
\end{center}
\vspace{5mm} 
We consider a general two-Higgs-doublet model with CP violation in the scalar sector,
that leads, at the one-loop level of the perturbation expansion,
to CP-violation in the process 
$\epem \to \ttbar \to l^\pm \cdots$ and $\epem \to \ttbar \to \bb \cdots$.
The goal of this study is to include {\it consistently} CP-violating effects
in distributions of top-quark decay products  ($l^\pm$ or $\bb$) that emerge {\it both} 
from $\ttbar$ production {\it and} from $t$ or $\bar{t}$ decay processes.

\vspace*{3cm}

PACS: 14.65.Ha, 14.80.Cp, 11.30.Er

\vspace*{.5cm}

Keywords: top quark, CP violation, two-Higgs-doublet model

\newpage
\renewcommand{\thefootnote}{\arabic{footnote}}
\pagestyle{plain} \setcounter{footnote}{0}  \setcounter{section}{0}


\section{Introduction}

Interactions of the top quark have not been precisely tested yet, in particular, 
CP violation in the top-quark interactions has not been verified. 
The classical method for
incorporating CP violation into the Standard Model (SM) of electroweak interactions
 is to make Yukawa couplings of the Higgs boson to
quarks explicitly complex,  as  built into the Kobayashi-Maskawa
mixing matrix~\cite{Kobayashi:1973fv} proposed more than two decades ago. 
However, CP violation could equally well be partially or wholly due
to other mechanisms. The possibility that 
CP violation derives largely from the Higgs sector 
itself is particularly appealing in the context of the observed baryon asymmetry,
since its explanation requires more CP violation~\cite{Gavela:1994dt,*Huet:1995jb} 
then is provided by the SM. Even 
the simple two-Higgs-doublet model  (2HDM) extension of the 
one-doublet SM Higgs sector
provides a much richer framework for describing CP violation since there 
spontaneous and/or explicit CP violation is possible
in the scalar sector~\cite{Lee:1973iz,*Weinberg:1990me,*Branco:1985aq}.
The model, besides CP violation, offers many other appealing phenomena, for 
a review see Ref.~\cite{HHG}. 

For our analysis, the most relevant part of the interaction Lagrangian 
takes the following form~\footnote{One could also consider more general, CP-violating $ZZh$ 
coupling, see Ref.~\cite{Han:2000mi}, however here the contribution from such a vertex
would be negligible.}:
\beq
{\cal{L}}= -\frac{\mt}{v}h\tbar(a+i\gamma_5 b)t + 
C \frac{h}{v} (\mz^2 Z_\mu Z^\mu+2 \mw^2 W_\mu W^\mu),
\label{lag}
\end{equation}
where $h$ is the lowest mass scalar, $g$ is the SU(2) coupling constant, 
$v$ is the Higgs boson vacuum expectation value 
(with the normalization adopted here such that $v=2m_W/g=246\,\gev$), 
$a$, $b$ and $C$ are real parameters which account for deviations from the
SM, $a=1$, $b=0$ and $C=1$ reproduce the SM Lagrangian. 
Since under CP, $\tbar(a+i\gamma_5 b)t \stackrel{{\rm CP}}{\ra} \tbar(a-i\gamma_5 b)t$ and 
$Z_\mu Z^\mu  \stackrel{{\rm CP}}{\ra} Z^\mu Z_\mu$, one can observe that 
terms in the cross section proportional to $ab$ or $bC$ would indicate 
CP violation. 
The~2HDM is the minimal extension of the SM that provides non-zero $ab$ and/or~$bC$.

In this paper we will focus on CP-violating contributions to the process 
$\epem \to \ttbar \to l^\pm  \cdots$ and $\epem \to \ttbar \to \bb \cdots$ 
induced within 2HDM. However the fundamental goal is seeking for the
ultimate theory of electroweak interactions. There are several reasons to utilize CP violation
in the top physics while looking for physics beyond the SM:
\bi
\item The top quark decays immediately after being produced as its huge
mass $m_t=174.0\pm3.2\pm4.0\gev$~\cite{Groom:2000in}
leads to a decay width ${\mit\Gamma}_t$ much
larger than~${\mit\Lambda}_{\rm QCD}$. 
Therefore the decay process is
not contaminated by any fragmentation 
effects~\cite{Bigi:1981az,*Kuhn:1982ua,*Bigi:1986jk} and decay
products may provide useful information on top-quark properties.
\item Since the top quark is heavy,
its Yukawa coupling is large and therefore its interactions could be 
sensitive to a Higgs sector of the electroweak theory. 
\item At the same time, the TESLA collider design is supposed to offer an integrated
luminosity of the order of $L=500\lumun$ at $\sqrt{s}=500\gev$. 
Therefore expected number of $\ttbar$ events per year could reach $5\times10^4$ 
even for $\ttbar$ tagging efficiency $\epsilon_{t\bar{t}}=15\%$. 
That should allow to study subtle properties of the top quark, which could
e.g. lead to CP-sensitive asymmetries of the order of~$5\times 10^{-3}$.
\item Since the top quark is that heavy  and the third family of quarks effectively 
decouples from the first two,
any CP-violating observables within the SM are expected to be 
tiny, e.g.: {\it i)} non-zero electric dipole 
moment of fermions is generated at the three-loop approximation of the perturbation 
expansion~\cite{Czarnecki:1997bu}, {\it ii)} the decay rate asymmetry (being a one-loop effect) 
is strongly GIM suppressed reaching at most a value $10^{-9}$~\cite{Grzadkowski:1993gh}.
So, one can expect that for CP-violating asymmetries any SM background could be safely
neglected.
\ei
Therefore it seems to be justified to look 
for CP-violating Higgs effects in the process of $\ttbar$ production and its subsequent decay at future
linear $\epem$ colliders. Even though 2HDM contributions to various CP-sensitive asymmetries
has been already published in the existing 
literature, see Refs.~\cite{Chang:1993fu,*Bernreuther:1992dz,Grzadkowski:1992yz}, 
here we are presenting results (for a detailed discussion see Ref.~\cite{bgjp01}) of a consistent
treatment of CP violation {\it both} in the production, $\epem \to \ttbar$,  {\it and} in the top-quark decay,
$t\to b W$.  For an extensive review of CP violation in top-quark interactions see 
Ref.~\cite{Atwood:2000tu}.

The paper is organized as follows. In Section~\ref{model}, we briefly outline the mechanism of CP 
violation in the 2HDM, introduce the mixing matrix for neutral scalars and derive necessary couplings.
In Section~\ref{expcon}, we recall current experimental constraints
relevant for the CP-violating observables considered in this paper.
In Section~\ref{asymm}, we collect results for the most attractive energy and angular 
CP-violating asymmetries.
Concluding remarks are given in Section~\ref{summary}.


\section{The two-Higgs-doublet model with CP \\violation}
\label{model}

The 2HDM of electroweak interactions 
 contains two SU(2) Higgs doublets denoted by 
$\Phi_1=(\phi_1^+,\phi_1^0)$ and $\Phi_2=(\phi_2^+,\phi_2^0)$.
It is well known~\cite{Lee:1973iz,*Weinberg:1990me,*Branco:1985aq} that the model 
allows both for spontaneous and explicit CP violation\footnote{Here we are considering
a model with discrete $Z_2$ symmetry that prohibits flavor changing neutral 
currents. In order to
allow for CP violation the symmetry has to be broken softly by the term 
$\mu_{12}^2\Phi_1^\dagger\Phi_2$ in the potential.}.

After
SU(2)$\times$U(1) gauge symmetry breaking, one combination of neutral
Higgs fields, $\sqrt2(\cb\Im\phi_1^0+ \sb\Im\phi_2^0)$,
becomes a would-be Goldstone boson which is absorbed while giving 
mass to the $Z$ gauge boson.
(Here, we use the notation $\sb\equiv\sin\beta$, $\cb\equiv\cos\beta$,
where $\tanb=\langle \phi_2^0 \rangle/\langle \phi_1^0 \rangle$.)
The same mixing angle, $\beta$, also diagonalizes 
the mass matrix in the charged Higgs sector.  
If either explicit or spontaneous CP violation is present,
the remaining three neutral degrees of freedom, 
\begin{equation}
(\varphi_1,\varphi_2,\varphi_3)\equiv
\sqrt 2(\Re\phi_1^0, \, \Re\phi_2^0, \,
 s_\beta\Im\phi_1^0-c_\beta\Im\phi_2^0) 
\end{equation} 
are not mass eigenstates. The physical neutral Higgs bosons $h_i$
($i=1,2,3$) are obtained by an orthogonal transformation, $h=R
\varphi$, where the rotation matrix is given in terms of three Euler
angles ($\alpha_1, \alpha_2,\alpha_3$) by
\begin{eqnarray} 
R=\left(\ba{ccc}
  c_1     &  -s_1c_2          &     s_1s_2  \\
  s_1c_3  & c_1c_2c_3-s_2s_3  &  -c_1s_2c_3-c_2s_3\\
  s_1s_3 & c_1c_2s_3+s_2c_3 & -c_1s_2s_3+c_2c_3 \ea\right),
\label{mixing}
\end{eqnarray}
where $s_i\equiv\sin\alpha_i$ and $c_i\equiv\cos\alpha_i$.

As a result of the mixing between real and imaginary parts of neutral
Higgs fields, the Yukawa interactions of the $h_i$ mass-eigenstates are not
invariant under CP. They are given by:
\begin{equation} 
{\cal L}=-\frac{m_f}{v} h_i\bar{f}(a^f_i+ib^f_i\gamma_5)f \label{coupl} 
\end{equation}
where the scalar ($a^f_i$) and pseudoscalar ($b^f_i$) couplings are
functions of the mixing angles. For up-type quarks we have 
\begin{equation}
a^u_i=\frac{1}{s_\beta}R_{i2},\;\;\;\;\;
b^u_i=\frac{c_\beta }{s_\beta}R_{i3}, 
\label{absu}
\end{equation}
and for down-type quarks:
\begin{equation}
a^d_i=\frac{1}{c_\beta}R_{i1},\;\;\;\;\;
b^d_i=\frac{s_\beta}{c_\beta} R_{i3}\,,
\label{absd}
\end{equation}
and similarly for charged leptons. 
For large $\tan\beta$, the couplings to down-type fermions are 
typically enhanced over the couplings to up-type fermions.

In the following analysis we will also need  
the couplings of neutral Higgs and vector bosons, they are given by
\begin{equation}
g_{VVh_i} \equiv 2 \frac{ m_V^2}{v} C_i= 
2 \frac{ m_V^2}{v} (s_{\beta} R_{i2}+c_{\beta}R_{i1}),
\label{zzcoups}
\end{equation}
for $V=Z,W$.
Hereafter we shall denote the lightest Higgs boson by $h$ and its $R$-matrix 
index by~$i$.


\section{Experimental Constraints}
\label{expcon}

Hereafter we will focus on  Higgs boson masses in the region, $m_h=10\div100\gev$.
As it has been shown in the literature~\cite{Gunion:1997aq}\cite{Abbiendi:2000ug} 
the existing LEP data are perfectly consistent
with one light\footnote{Sum rules discussed in Ref.~\cite{Gunion:1997aq} prove that
even within the CP-violating version of the 2HDM one can satisfy LEP 
experimental constraints with one light Higgs boson.}
Higgs boson within the 2HDM. It turns out that even precision electroweak
tests allow for light Higgs bosons~\cite{Chankowski:1999ta}.

In order to amplify the form factors calculated in this paper we have adopted for
an illustration $\tanb =0.5$. However, there exist experimental constraints on $\tanb$ from
${\rm K^0 - \bar{K}^0}$ and ${\rm B_d-\bar{B}_d}$ mixing~\cite{Gunion:1990vk},
$b\to s \gamma$ decay~\cite{Greub:1999sv} and $Z\to b\bar{b}$ decay~\cite{Haber:1999zh}. 
Since small $\tanb$ enhances $H^\pm t b$ coupling, 
in order to maintain $\tanb =0.5$ we have to decouple charged Higgs effects and therefore
we assume that $m_{H^\pm} \gsim 500\div600\gev$.

The constraints on the mixing angles 
$\alpha_i$  that should be imposed in our numerical analysis are as follows:
\begin{itemize}
\item The $ZZh$ couplings, $C_i^2$,  are restricted by non-observation of
Higgs-strahlung events at LEP1 and LEP2, see Ref.~\cite{Abbiendi:1998rd,*ALEPH2000-028}
\item The contribution to the total $Z$-width from $Z \ra Z^* h_i \ra f
\bar{f} h_i$  is required to be below $7.1 \mev$, see
Ref.~\cite{Ackerstaff:1998ms}.
\end{itemize}
It turns out that the restriction on the $ZZh$ coupling from its contribution to 
the total $Z$-width is always weaker then the one from $Zh$ production if
$m_h \gsim 10\gev$.

The LEP constraints on the $ZZh$ coupling restrict the following entries of the mixing matrix $R_{ij}$:
\begin{equation}
|\sin\beta R_{i2} + \cos_{\beta} R_{i1}|\leq C_i^{exp},
\label{eq:2hdm:LEPZZh}
\end{equation}
where $C_i^{exp}$ stands for the upper limit for the relative strength of $ZZh$ coupling  
determined experimentally in Ref.~\cite{Abbiendi:1998rd,*ALEPH2000-028} up to the Higgs mass $m_h=105\gev$.
As we have concluded in the previous section, CP-violating phenomena we are considering
are enhanced by small $\tan \beta$, in that case one can see from 
Eq.(\ref{eq:2hdm:LEPZZh}) that the LEP constraints mostly
restrict $R_{i1}$. Through the orthogonality the restriction on $R_{i1}$  is being transfered to constrain  
$|R_{i2}R_{i3}|=|R_{i2}\sqrt{1-R_{i1}^2-R_{i2}^2}|$ which 
multiplies leading contributions to all CP-violating asymmetries considered here.\footnote{As it has been shown 
in Ref.~\cite{bgjp01} the other contribution that is 
proportional to $R_{i1}R_{i3}$ is by 1-2 orders of magnitude smaller.} 
The final result for the upper limit on $|R_{i2}R_{i3}|$ 
as a function of $\tan \beta$ is shown in Fig.\ref{fig:2hdm:ogrLEPtanb}. In fact the bound on 
$|R_{i2}R_{i3}|$ depends on the Higgs mass, however, in order to be conservative, we have
assumed  $C_i^{exp}=0.12$ that is the most restrictive experimental limit 
(obtained for $\mh\simeq 18\gev$\footnote{ 
For $\mh\simeq 18\gev$ the limits presented 
in Fig.16 of Ref.~\cite{Abbiendi:1998rd,*ALEPH2000-028} 
for  the case when no $b$-tagging  and with $b$-tagging almost coincide.
Therefore our plot in Fig.\ref{fig:2hdm:ogrLEPtanb} is not influenced by
potential problems concerning the dependence of the 
Higgs-$\bbbar$ and Higgs-$\tau^+\tau^-$ branching ratios on the mixing
angles.}).

\jpfigure{width=0.65\textwidth}{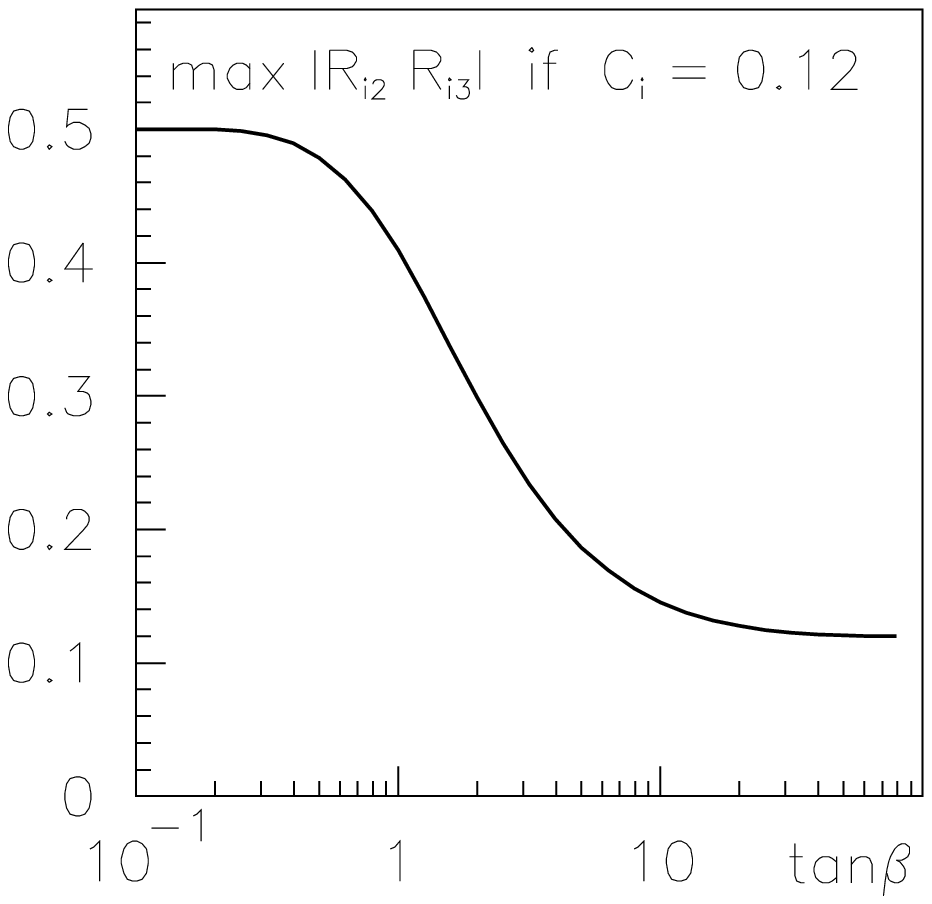}{Maximal value of 
 $|R_{i2}R_{i3}|$ allowed by the LEP constraints on $ZZh_i$ coupling
as a function of $\tb$.}{fig:2hdm:ogrLEPtanb}
 
As it is seen from Fig.\ref{fig:2hdm:ogrLEPtanb} the constraints for 
$|R_{i2}R_{i3}|$
are weak for small $\tanb$. Therefore for $\tan\beta\simeq 0.5 $
it should be legitimate to assume $|R_{i2}R_{i3}|\simeq 1/2$ which is the maximal value consistent 
with orthogonality.


\section{CP-Violating Asymmetries}
\label{asymm}
Hereafter we assume that there exists only one light Higgs boson $h$ and 
possible effects of 
the heavier scalar degrees of freedom decouple.

The effective $\ttbar \gamma$ and $\ttbar Z$  vertices will be parameterized by the following form 
factors\footnote{Two other possible
    form factors do not contribute in the limit of zero electron 
    mass.}:

\begin{equation}
{\mit\Gamma}^\mu_v=\frac{g}{2}\bar{u}(p_t)
\biggl[\,\gamma^\mu(A_v-B_v\gamma_5)
+\frac{(p_t-p_{\bar{t}})^\mu}{2\mt}(C_v-D_v\gamma_5)\,\biggr]v(p_{\bar{t}}),
\label{vtt}
\end{equation}
where $g$ denotes the SU(2) gauge coupling constant and $v=\gamma,Z$.
The SM contributions to the form factors
are the following:
$$
A_\gamma^{(SM)}=-\frac43\sw,\ \ B_\gamma=0,\ \ 
A_Z^{(SM)}=-\frac{v_t}{2\cw},\ \ 
B_Z^{(SM)}=-\frac{a_t}{2\cw}
$$
for
$$
v_t=\Bigl(1-\frac83\sin^2\theta_W\Bigr) \lsp
a_t=1.
$$
The 
form factors $A_v$, $B_v$, $C_v$ describe
$CP$-conserving while $D_v$ parameterizes $CP$-violating
contributions. 

Further in this paper the following parameters will be adopted:
$m_t=175\gev$, $m_Z=91.187\gev$, 
$\Gamma_Z=2.49\gev$, $\sin^2 \theta_W=0.23$ and $m_b=4.2\gev$.

Direct calculation of appropriate Feynman diagrams  leads to 
the following result~\cite{bgjp01} in terms of 3-point Passarino-Veltman~\cite{PassarinoVeltman}  
functions defined in the appendix of Ref.~\cite{bgjp01}:
\bea
\label{eqn:eett:dgammadz}
D_\gamma&=&
\frac{\ri}{2\pi^2} A_\gamma
\frac{m_t^2}{v^2} 
b_i^t a_i^t  m_t^2 
\loopC{12}{}(p_t, p_{\bar{t}}, m_t^2, m_h^2, m_t^2),\nnr\\
D_Z&=& 
\frac{\ri}{2\pi^2} A_Z  \frac{m_t^2}{v^2}
b_i^t \left[
a_i^t m_t^2 
\loopC{12}{}(p_t,  p_{\bar{t}}, m_t^2, m_h^2, m_t^2)\right.\nnr\\
&& 
\left. -  
C_i m_Z^2 
\loopC{12}{}(p_t,p_{\bar{t}}, m_h^2, m_t^2, m_Z^2) \right].
\eea
From Eq.(\ref{eqn:eett:dgammadz}) and  Eqs.(\ref{absu}, \ref{zzcoups}) one can find out 
that all contributions to the form factors $D_{\gamma}$, $D_Z$ are enhanced for  small $\tanb$.

We will adopt the following parameterization of
the $Wtb$ vertex suitable for the $t$ and $\tbar$ decays:
\bea
{\Gamma}^{\mu}&=&-\frac{g}{\sqrt{2}}V_{tb}\:
\biggl[\,\gamma^{\mu}(f_1^L P_L +f_1^R P_R)
-\frac{{i\sigma^{\mu\nu}k_{\nu}}}{M_W}
(f_2^L P_L +f_2^R P_R)\,\biggr],\ \ \ \ \non \\
\bar{\Gamma}^{\mu}&=&-\frac{g}{\sqrt{2}}V_{tb}^*\:
\biggl[\,\gamma^{\mu}(\bar{f}_1^L P_L +\bar{f}_1^R P_R)
-\frac{{i\sigma^{\mu\nu}k_{\nu}}}{M_W}
(\bar{f}_2^L P_L +\bar{f}_2^R P_R)\,\biggr],
\label{tbw}
\eea
where $P_{L/R}=(1\mp\gamma_5)/2$, $V_{tb}$ is the $(tb)$ element of
the Kobayashi-Maskawa matrix and $k$ is the momentum of $W$. 
In the SM $f_1^L=\bar{f}_1^L=1$ and all the other form factors vanish.
It turns out that in the limit of massless bottom quarks the only form factors that interfere with
the SM are $f_2^R$ and $\bar{f}_2^L$ for the top and anti-top decays, respectively.
Currently, there is no relevant 
experimental bound on those form factors\footnote{There exists however 
direct experimental constraint from the Fermilab Tevatron on the form factor $f_1^R$,
that are obtained through the determination of the $W$-boson helicity. Pure $V-A$
theory for massless bottom quarks predicts an absence of positive helicity
$W^+$ bosons, therefore the upper limit on the helicity ${\cal F}_+$ implies
an upper limit on the $V+A$ coupling $f_1^R$, however, the resulting  
limit is rather weak~\cite{Affolder:1999mp}. There exist an indirect, but much stronger 
bound~\cite{Cho:1994zb,*Fujikawa:1994zu} on
the admixture of right-handed currents, $\bar{f}_1^R$, coming from data for 
$b \to s \gamma$, namely $-0.05\lsim\bar{f}_1^R\lsim0.01$.}.

One can show that the CP-violating and CP-conserving parts of the
form factors for $t$ and $\tbar$ are not independent:
\beq
f_1^{L,R}=\pm\bar{f}_1^{L,R} \;\;{\rm and} \;\;\;f_2^{L,R}=\pm\bar{f}_2^{R,L},
\label{cp_relation}
\end{equation}
where upper (lower) signs are those for $CP$-conserving
(-violating) contributions~\cite{Bernreuther:1992be,Grzadkowski:1992yz}. Therefore any
$CP$-violating observable defined for the top-quark decay must be
proportional to $f_1^{L,R}-\bar{f}_1^{L,R}$ or $f_2^{L,R}-
\bar{f}_2^{R,L}$.

At the one-loop level one gets the following
result~\cite{bgjp01}  for CP-violating contribution to $\Re(f_2^R|_{CPV})$:
\begin{equation}
\Re(f_2^R-\bar{f}_2^L)=2\Re(f_2^R|_{CPV})=\frac{g }{16 \pi^2 } \frac{m_b}{v} m_b m_t
    b_i^b C_i \Im \loopC{22}{bd}
\label{f2r}
\end{equation}

Adopting the maximal value of $R_{i2}R_{i3}$ allowed by the orthogonality and the LEP constraints
for  $\tanb=0.5$, we may discuss a possibility for an 
experimental determination of the calculated form 
factors at future $\epem$ colliders.
A detailed discussion of expected statistical uncertainties for a measurement of the form factors
has been performed in  Ref.~\cite{Grzadkowski:2000nx}. It has been shown that adjusting
an optimal $\epem$ beam polarizations,
using the energy and angular double distribution of final leptons 
and fitting {\it all 9 form factors} leads
to the following statistical errors for the determination of CP-violating
form factors: $\Delta[\Re(D_\gamma)]=0.08$ and
$\Delta[\Re(D_Z)]=14.4$ for $\epsilon_{\ttbar}\simeq 15\%$.
It is seen that only $\Re(D_\gamma)$, could be
measured with a high precision. We have found (see plots in Ref.~\cite{bgjp01}) 
that $\Re(D_\gamma)$ may reach at most a value of $0.10$, therefore one shall conclude that
several years of running with yearly integrated luminosity $L=500\lumun$ should allow for an observation
of $\Re(D_\gamma)$ generated within 2HDM, provided the lightest Higgs boson mass is not too large.
On the other hand, the expected~\cite{Grzadkowski:2000nx} precision for the determination 
of the decay form factors is much more promising:  
$\Delta[\Re(f_2^R-\bar{f}_2^L)]=0.014$. However, it has been found in Ref.~\cite{bgjp01} that  
the maximal expected\footnote{It turns out (see Ref.~\cite{bgjp01} for details) that
$\Re(f_2^R-\bar{f}_2^L)$ is by $2-4$ orders of magnitude below $\Re D_\gamma$
or $\Re D_Z$ even for large b-quark Yukawa coupling ($\tanb = 50$).
The~suppression is caused both by the experimental limit on 
$|C_i|$  (for $m_h < 105\gev$)
and by an extra suppression factor of $(m_b/m_t)^2$ 
(relative to $\Re D_{\gamma,Z}$).}  
size of $\Re(f_2^R-\bar{f}_2^L)$
is  $5 \times 10^{-5}$ (for $m_h > 10\gev$),
therefore either an unrealistic growth of the luminosity, or other observables (besides the energy
and angular double distribution of final leptons)  are required in order to 
observe CP-violating from factors in the top-quark decay process. The results
of Ref.~\cite{Grzadkowski:2000nx} assumed simultaneous\footnote{Obviously, that leads to reduced
precision for the determination of the form factors.} determination of  {\it all 9 form factors}, therefore
another chance to reduce
of $\Delta[\Re(f_2^R-\bar{f}_2^L)]$ is to have some extra independent constraints on the top-quark
coupling coming from other colliders, like the Fermilab Tevatron or LHC.

Looking for CP violation one can directly measure in the model independent 
way~\cite{Grzadkowski:2000nx}  
all the form factors including those which are odd under CP. However another possible 
attitude is to construct  certain asymmetries sensitive to CP violation.
In this section we will discuss several asymmetries that could probe CP violation in the processes
$\epem \to \ttbar \to l^\pm \cdots$ and $\epem \to \ttbar \to \bb \cdots$.
We will systematically drop all contributions quadratic in 
non-standard form factors and
calculate various asymmetries keeping only interference between the SM and $D_\gamma$, $D_Z$ 
or $\Re(f_2^R-\bar{f}_2^L)$.

\subsection{Integrated Lepton-Energy Asymmetry}
Let us introduce the rescaled lepton energy, $x$, by
\beq
x\equiv
\frac{2 E_l}{\mt}\left(\frac{1-\beta_t}{1+\beta_t}\right)^{1/2},
\label{def-x}
\end{equation}
where $E_l$ is the energy of $l$ in $\epem$ c.m. frame and $\beta_t\equiv \sqrt{1-4m_t^2/s}$.

CP symmetry could be tested using the following 
leptonic double energy distribution~\cite{Grzadkowski:1997pc}:
\beq
\frac{1}{\sigma}\frac{d^2\sigma}{dx\;d\barx}
=\sum_{i=1}^{3}c_i f_i(x,\barx),
\label{DD}
\end{equation}
where $x$ and $\barx$ are for $l^+$ and $l^-$, respectively, and
$$
c_1=1,\ \ \ 
c_2=\xi,\ \ \ 
c_3=\frac{1}{2}\Re(f_2^R-\bar{f}_2^L)
$$
for
\begin{eqnarray*}
&&\xi \equiv\frac{1}{(3-\beta^2)D_V+2\beta^2 D_A} \\
&&\ \ \ \ \ \times
\frac{-1}{\sin\theta_W}{\Re}\biggl[\:\frac23\: D_\gamma
+\frac{s^2}{(s-m_Z^2)^2}
\frac{(v_e^2+a_e^2)v_t}{64\sin^3\theta_W\cos^3\theta_W}D_Z
\\
&&\ \ \ \ \
-\frac{s}{s-m_Z^2}
\biggl(\,\frac{v_e v_t}{16\sin^2\theta_W\cos^2\theta_W}D_\gamma
+\frac{v_e}{6\sin\theta_W\cos\theta_W}D_Z \,\biggr)\:\biggr],
\end{eqnarray*}
for
\begin{eqnarray*}
&&D_V=(v_e v_t d-\frac23)^2 +(a_e v_t d)^2, \\
&&D_A=(v_e a_t d)^2 +(a_e a_t d)^2,         \\
\end{eqnarray*}
with the SM neutral-current parameters of $e$: 
$v_e=-1+4\sin^2\theta_W$, $a_e=-1$ and a $Z$-propagator factor
$$
d\equiv\frac{s}{s-m_Z^2}
\frac{1}{16\sin^2\theta_W\cos^2\theta_W}.
$$
The definitions of the functions $f_i(x,\bar{x})$ are to be found in
Ref.~\cite{Grzadkowski:1997kn}.

The coefficients $c_2$ and $c_3$ measure the degree of CP violation in the $\ttbar$ production and
their subsequent decays, respectively. The following asymmetry could be defined~\cite{bgjp01} to extract
$\ReDgamma$, $\ReDZ$ and $\Refrmfl$
form the double energy distribution:
\begin{equation}
A_{CP}^{ll}\equiv
\frac
{\dps\int\int_{x<\bar{x}}dxd\bar{x}\frac{d^2\sigma}{\dps dxd\bar{x}}
 -\int\int_{x>\bar{x}}dxd\bar{x}\frac{d^2\sigma}{\dps dxd\bar{x}}}
{\dps\int\int_{x<\bar{x}}dxd\bar{x}\frac{d^2\sigma}{\dps dxd\bar{x}}
 +\int\int_{x>\bar{x}}dxd\bar{x}\frac{d^2\sigma}{\dps dxd\bar{x}}}.
\label{asy_9608}
\end{equation}
In order to estimate a relative strength of various sources\footnote{It should be noticed that
the general formulae (see 
Refs.~\cite{Grzadkowski:2000nx},\cite{Grzadkowski:1997kn},\cite{Grzadkowski:1997pc},\cite{Grzadkowski:1999iq})
for the asymmetries considered here have been obtained assuming $\mb=0$. As it is seen from
Eq.(\ref{f2r}),  the contribution to CP violation in the decay process, $\Refrmfl$, turns out
to be proportional to $\mb^2$. Therefore, strictly speaking, CP violation in the decay process
should either be disregarded or all the CP-violating contributions of the order of $\mb^2$ 
should be calculated. The latter effects are definitely negligible  in the 2HDM comparing to
contributions from the production process.
However, we have found it useful for future applications within other
possible models\cite{work_in_progress}
to preserve hereafter contributions  from $\Refrmfl$ in formulae and corresponding figures
for all the asymmetries considered in this study.}
of CP violation it is worth to decompose the asymmetry as follows:
\begin{equation}
A_{CP}^{ll}=\ggttbar^{ll}\ \ReDgamma+
            \gZttbar^{ll}\ \ReDZ+
            \gWtb^{ll}   \ \Refrmfl.
\label{h_def}
\end{equation}
In Table~\ref{s_dep_ff} we show the coefficients $g$ for various c.m. energies.
Firstly, is clear that for any given $\sqrt{s}$ the coefficient $\gZttbar^{ll}$ is the smallest one.
Secondly, it is seen that just above the threshold for $\ttbar$ production there is an enhancement of
relative contributions from the decay, however that still not sufficient to overcome the suppression
of $\Re(f^R_2-\bar{f}^L_2) $.
Therefore we can conclude  that the leading contribution is provided by CP violation in 
the $\gamma\ttbar$ vertex. 

\vspace{1cm}
\begin{table}[h]
\begin{center}
\begin{tabular}{|r|r @{.} l|r @{.} l|r @{.} l|}
\hline
$\sqrt{s}$[GeV]& 
\multicolumn{2}{c|}{$\ggttbar^{ll}$} & 
\multicolumn{2}{c|}{$\gZttbar^{ll}$}&
\multicolumn{2}{c|}{$\gWtb^{ll}$}\\
\hline
360&0&0509 & 0&00954 & 0&410 \\
\hline
500&0&386 & 0&0684 & 0&291 \\
\hline
1000&0&602 & 0&102 & 0&235 \\
\hline
\end{tabular}
\end{center}
\caption{The energy dependence of the coefficients $g$ defined in Eq.(\ref{h_def}).}
\label{s_dep_ff}
\vspace*{0.5cm}
\end{table}

Fig.\ref{fig:eett:asy9608r:mhdep} illustrates the Higgs-mass dependence of the leading
(proportional to $R_{i2} R_{i3}$) contribution to the integrated lepton-energy asymmetry.
It turns out that $\sqrt{s}=500\gev$ provides the largest asymmetry.

\jpfigure{width=.65\textwidth}{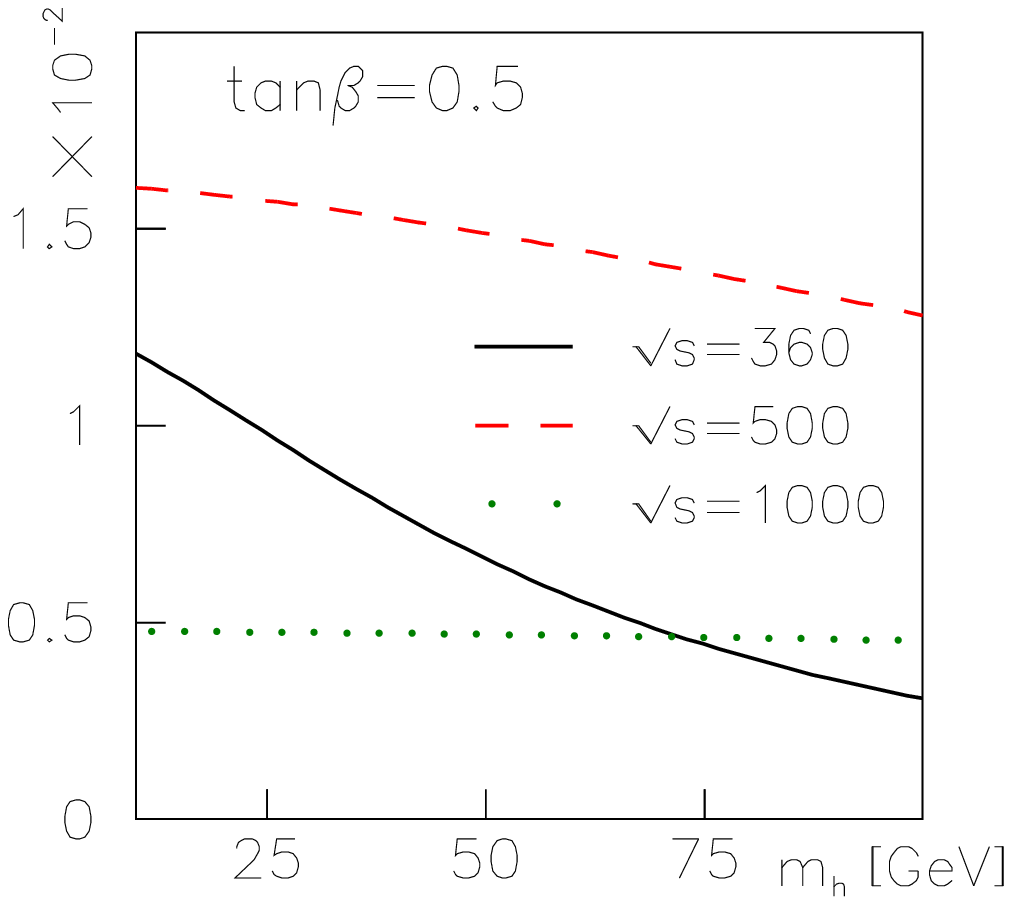}{
Higgs mass dependence of the coefficient of
$R_{i2} R_{i3}$  for the asymmetry given by Eq.(\ref{asy_9608})
for $\sqrt{s}$=360 (solid), 500 (dashed), 1000 GeV (dotted) for 
$\tanb=0.5$. }{fig:eett:asy9608r:mhdep}

Using results of Ref.~\cite{Grzadkowski:1997pc} one can find out an expected
statistical error for the determination of $A_{CP}^{ll}$ at any given $\epem$ collider.
Assuming $\sqrt{s}=500\gev$, $L=500\lumun$ and lepton tagging efficiency, 
$\epsilon_l=60\%$ we get $\Delta A_{CP}^{ll}=0.014$. 
As it is seen from Fig.\ref{fig:eett:asy9608r:mhdep}
an observation of the asymmetry would require several years of running
at the assumed luminosity.

\subsection{Integrated Angular Asymmetry}

Another CP-violating asymmetry could be constructed using
the angular distributions of the bottom quarks or leptons originating from 
the top-quark decay:
\beq
\frac{d\sigma}{d\cos\thf}=
\frac{3\pi\beta\alpha_{\mbox{\tiny EM}}^2}{2s}B_{\sst{f}}
\left({\mit\Omega}_0^{\sst{f}}+{\mit\Omega}_1^{\sst{f}}
\cos\thf+{\mit\Omega}_2^{\sst{f}}\cos^2\thf\right),   
\label{dis2}
\end{equation}
where $f=b,l$, $B_f$ is an appropriate 
top-quark branching ratio, $\theta_f$ is the angle between the $e^-$ beam direction and the
direction of $f$ momentum in the $\epem$ c.m. frame and $\Omega_i^f$ are coefficients
calculable in terms of the form factors, see Ref.~\cite{Grzadkowski:1999iq}.
The following asymmetry provides a signal of CP violation:
\begin{equation}
{\cal A}_{\sst{CP}}^{\ssf}(P_{e^-},P_{e^+})= \frac{
{\displaystyle \int_{-c_m}^{0}\!d\cos\thf
 \frac{d\sigma^{+(*)}(\thf)}{d\cos\thf}
 -\int_{0}^{+c_m}\!d\cos\thf \frac{d\sigma^{-(*)}(\thf)}{d\cos\thf}}}
{{\displaystyle \int_{-c_m}^{0}\!d\cos\thf
  \frac{d\sigma^{+(*)}(\thf)}{d\cos\thf}
 +\int_{0}^{+c_m}\!d\cos\thf \frac{d\sigma^{-(*)}(\thf)}{d\cos\thf}}}, 
\label{asy_0004}
\end{equation}
where $P_{e^-}$ and $P_{e^+}$ are the polarizations of $e$ and
$\bar{e}$ beams, $d\sigma^{+/-(*)}$ is referring to $f$ and $\bar{f}$
distributions respectively, and $c_m$ expresses the experimental
polar-angle cut. In order to discuss possible advantages of polarized initial beams
we are considering here dependence of the asymmetry on
the polarization. Hereafter we will discuss the same polarization for
$e$ and $\bar{e}$: $P\equiv P_{e^-}=P_{e^+}$. 
 
Again we decompose the asymmetry as follows:
\begin{equation}
{\cal A}_{CP}^f(P)=\ggttbar^f(P)\ \ReDgamma+ 
                   \gZttbar^f(P)\ \ReDZ+
                   \gWtb^f(P)   \ \Refrmfl.
\label{g_def}
\end{equation}
In Table~\ref{s_dep_g} we show the coefficient functions $g$ calculated for various energy
and polarization choices assuming the polar angle cut $|\cos\thf|<
0.9$, i.e. $c_m=0.9$ in Eq.(\ref{asy_0004}), both for leptons and bottom
quarks\footnote{Note that in Table~\ref{s_dep_g} there is no column corresponding to the coefficient of 
$\Re(f^R_2-\bar{f}^L_2)$. That happens since the angular distribution for leptons is not 
influenced by corrections to the top-quark decay vertex, see 
Refs.~\cite{Grzadkowski:1999iq,Rindani:2000jg} 
and~\cite{Grzadkowski:2000nx}.} . It could be seen that a positive polarization leads to 
higher coefficients $g_{\gamma\ttbar}^f$ and
$g_{Z\ttbar}^f$.
Since $\Re(D_\gamma) > \Re(D_Z) \gg \Re(f^R_2-\bar{f}^L_2) $ that implies that 
maximal asymmetry could be reached for $P=+0.8$ and 
the dominant contribution is originating from $\Re(D_\gamma)$.
Since the number of events does not drop drastically when going from unpolarized beams to $P=+0.8$,
it turns out that the positive polarization is the most suitable for testing the integrated angular asymmetry.
It is clear from the table that the asymmetry for final leptons should be larger by a factor $3\div4$ than
the one for bottom quarks and their signs should be reversed.

\begin{table}
\begin{center}
\begin{tabular}{|r|r|r @{.} l|r @{.} l|r @{.} l|r @{.} l|r @{.} l|}
\hline
$\sqrt{s}[\gev]$&$P$&\multicolumn{6}{c|}{quark b}&\multicolumn{4}{c|}{lepton}\\
\cline{3-12}
&&
\multicolumn{2}{c|}{$g_{\gamma\ttbar}^b(P)$} & 
\multicolumn{2}{c|}{$g_{Z\ttbar}^b(P)$}&
\multicolumn{2}{c|}{$g_{Wtb}^b(P)$}&
\multicolumn{2}{c|}{$g_{\gamma\ttbar}^l(P)$} & 
\multicolumn{2}{c|}{$g_{Z\ttbar}^l(P)$}\\ 
\hline
360 & 0.0& 0&00844& 0&00106& 0&142& -0&0162&-0&00203\\ 
    & 0.8& 0&00983&-0&00555&-0&259& -0&0493& 0&0278\\
    &-0.8& 0&00758& 0&00510& 0&388& -0&0106&-0&00713\\
\hline 
500 & 0.0& 0&113&   0&0136&  0&121& -0&224& -0&0270\\
    & 0.8& 0&131&  -0&0718& -0&247& -0&627&  0&343\\
    &-0.8& 0&101&   0&0661&  0&347& -0&149& -0&0968\\
\hline
1000& 0.0& 0&332&   0&0389&  0&0678&-0&722& -0&0845\\
    & 0.8& 0&422&  -0&225&  -0&167& -1&55&   0&824\\
    &-0.8& 0&284&   0&181&   0&194& -0&507& -0&322\\
\hline
\end{tabular} 
\end{center}
\caption{The energy and polarization dependence of the coefficients $g_{\gamma\ttbar}^f(P)$,
$g_{Z\ttbar}^f(P)$ and  $g_{Wtb}^f(P)$ defined in Eq.(\ref{g_def}) for leptons ($f=l$) 
and bottom quarks ($f=b$).}
\label{s_dep_g}
\end{table}

Using the general formula for the asymmetry 
from Ref.~\cite{Grzadkowski:2000nx} and adopting results for the CP-violating form factors
we plot ${\cal A}_{\sst{CP}}^{\ssf}(P_{e^-},P_{e^+})$ in Fig.\ref{fig:eett:asy0004br-mhdep}
as a function of the Higgs mass both for bottom quarks and leptons. It is clear that the largest 
asymmetry could be expected for $P_{e^-}=P_{e^+}=+0.8$ for final leptons at $\sqrt{s}=500\gev$.
With the maximal mixing, $R_{i2}R_{i3}=1/2$ the $1\%$ asymmetry could be expected for
the Higgs boson with mass $m_h=10\div20\gev$. Since the statistical error 
expected~\cite{Grzadkowski:2000nx} for
the asymmetry is of the order of $5\times 10^{-3}$, we can conclude that the asymmetry
${\cal A}_{\sst{CP}}^{\ssf}(P_{e^-},P_{e^+})$ is the most promising one, leading to
$2\sigma$ effect for light Higgs mass and $\tanb=0.5$. 
As it is seen form Fig.\ref{fig:eett:asy0004br-mhdep}
it is relevant to have polarized $\epem$ beams.

\jpfigure{width=.65\textwidth}{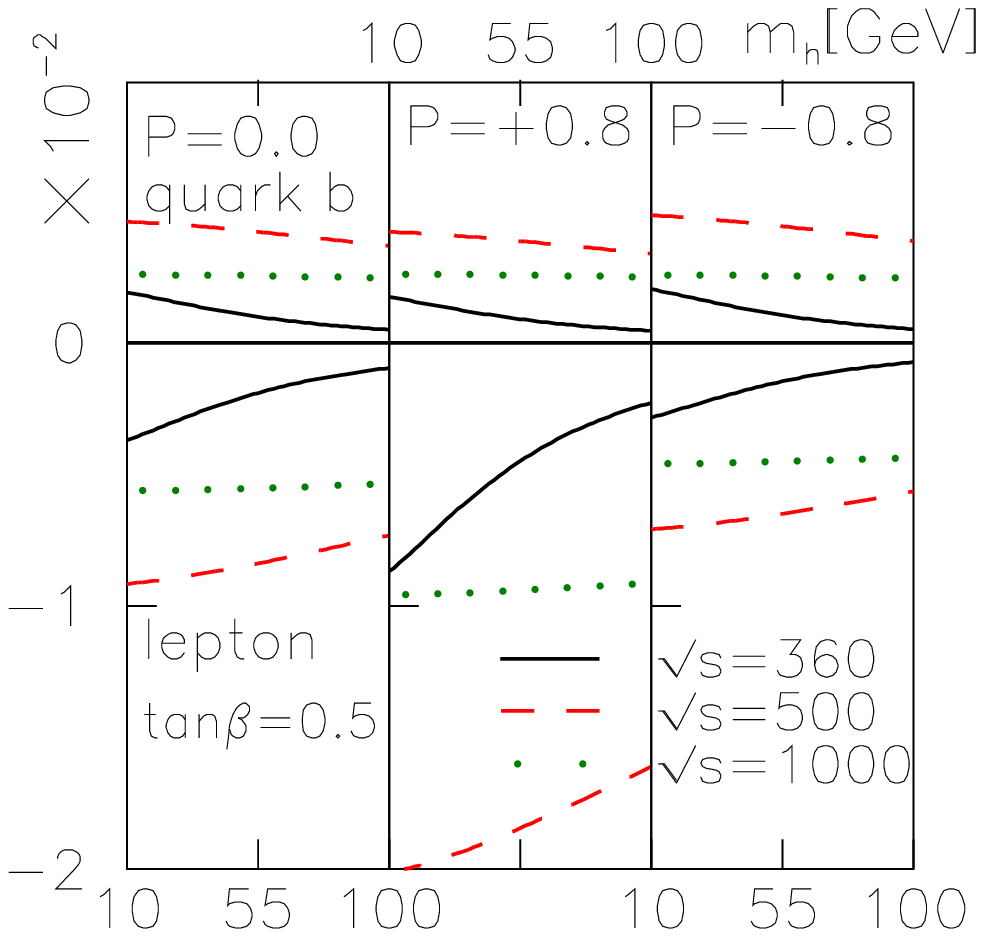}{The Higgs mass dependence of 
the coefficient  of $R_{i2} R_{i3}$ for the angular asymmetry defined by Eq.(\ref{asy_0004})
for bottom quarks (upper) and leptons (lower) at $\sqrt{s}$=360 (solid), 500 (dashed), 
1000 GeV (dotted) with unpolarized beams (left), $P=+0.8$ (middle) and $P=-0.8$ (right)
for $\tanb=0.5$.}{fig:eett:asy0004br-mhdep}


\section{Summary}
\label{summary}

We have considered a general two-Higgs-doublet model with CP violation in the scalar sector.
Mixing of the three neutral Higgs fields of the model leads to CP-violating Yukawa couplings
of the physical Higgs bosons. CP-asymmetric form factors generated at the one-loop level of 
perturbation theory has been calculated within the model.
Although in general the existing experimental data from LEP1 and LEP2 constraint
the mixing angles of  the three neutral Higgs fields, their combination relevant
for CP violation is not bounded for small $\tanb$ which is the region of our interest.
We have shown that the decay form factors are typically smaller then the production ones
by 2-3 orders of magnitude. The dominant contribution to CP violation in the production is coming from
$\gamma\ttbar$ coupling. Several energy and angular
CP-violating asymmetries for the process
$\epem \to \ttbar \to l^\pm \cdots$ and $\epem \to \ttbar \to \bb \cdots$ has been considered
using the form factors calculated within the two-Higgs-doublet model.
It turned out that the best test of CP invariance would be provided by the integrated 
angular asymmetry ${\cal A}_{\sst{CP}}^{\ssf}(P_{e^-},P_{e^+})$ for positive 
polarizations of $\epem$ beams. For one year of running at TESLA collider with the integrated luminosity
$L=500\lumun$ one could  expect $2\sigma$ effect  for the asymmetry for light Higgs boson
and $\tanb=0.5$.

\vspace{.5cm}
\centerline{\bf Acknowledgments}
\vspace{.5cm} This work was supported in part by the Committee for
Scientific Research (Poland) under grants No.~5~P03B~121~20 and No.~2~P03B~080~19.

\bibliographystyle{phcpc_mcite}

\end{document}